**Thickness Dependent OER Electrocatalysis of Epitaxial $LaFeO_3$ Thin Films**

Andricus R. Burton[a†], Rajendra Paudel[b†], Bethany Matthews[c], Michel Sassi[d], Steven R. Spurgeon[c], Byron H. Farnum[a]*, Ryan B. Comes[b]*

[a]Department of Chemistry and Biochemistry, Auburn University, Auburn, AL 36849

[b]Department of Physics, Auburn University, Auburn, AL 36849

[c]Energy and Environment Directorate, Pacific Northwest National Laboratory, Richland, WA 99352

[d]Physical and Computational Sciences Directorate, Pacific Northwest National Laboratory, Richland, WA 99352

*corresponding authors: farnum@auburn.edu, ryan.comes@auburn.edu

†equal contribution

**Abstract**

Transition metal oxides have long been an area of interest for water electrocatalysis through the oxygen evolution and oxygen reduction reactions. Iron oxides, such as $LaFeO_3$, are particularly promising due to the favorable energy alignment of the valence and conduction bands comprised of $Fe^{3+}$ cations and the visible light band gap of such materials. In this work, we examine the role of band alignment on the electrocatalytic oxygen evolution reaction (OER) in the intrinsic semiconductor $LaFeO_3$ by growing epitaxial films of varying thicknesses on Nb-doped $SrTiO_3$. Using cyclic voltammetry and electrochemical impedance spectroscopy, we find that there is a strong thickness dependence on the efficiency of electrocatalysis for OER. These measurements are understood based on interfacial band alignment in the system as confirmed by layer-resolved electron energy loss spectroscopy and electrochemical Mott-Schottky measurements. Our results demonstrate the importance of band engineering for the rational design of thin film electrocatalysts for renewable energy sources.

## Introduction

Energy storage via water splitting is a central goal toward the realization of a renewable energy economy. Formally, this process requires the oxidation of water to dioxygen ($O_2$) and reduction of protons to dihydrogen ($H_2$) whereby 4.92 eV can be stored per $H_2$ molecule (**Equation 1**). The anodic half-reaction, termed the oxygen evolution reaction (OER), is the more challenging step as it requires multiple protons and oxidizing equivalents in addition to the formation of an O=O double bond (**Equation 2**). Among the many catalysts which have been studied to promote this reaction, transition metal oxides remain an interesting target given their relative ease of synthesis, aqueous stability (depending on pH), wide range in crystal structures, and the diversity of available transition metal cations.[1–4] More importantly, the use of first-row transition metals, which are readily abundant in the Earth's crust, make these oxides attractive from a cost perspective as a means of circumventing expensive metals such Pt, Ru, and Ir as catalytic materials.

1) $2H_2O \rightarrow 2H_2 + O_2$ $\quad \Delta G^o = 4.92$ eV

2) $2H_2O \rightarrow O_2 + 4H^+ + 4e^-$ $\quad E^o = 1.23$ V vs RHE

Perovskite oxides, with the formula $ABO_3$, have gained particular attention toward OER electrocatalysis since the seminal publication by Suntivich *et al* in which the overpotential for catalysis was shown to vary with the number of $e_g$ electrons present on the B-site metal according to the Sabatier principle.[5] An optimal number of $e_g$ electrons was found to be ~1.1. The catalytic performance of perovskite oxides have also been found to depend significantly on several other factors such as surface termination[6,7] and epitaxial strain[8,9], with single crystal thin films demonstrating catalytic performance that matches that of Pt for OER.[8,10]

Band alignment at the interface between single crystal thin films and their substrates is also critical to their catalytic performance. In particular, the $SrTiO_3/LaFeO_3$ (STO/LFO) interface has attracted interest for questions of band engineering for catalytic applications.[11–15] To examine these interfaces, we used molecular beam epitaxy (MBE) to synthesize epitaxial single crystal films with varying thicknesses from 2 – 10 nm. We show that the electrocatalytic current for OER is greatly dependent on thickness where a sharp increase in current between 2 and 5 nm was observed while current decreased over the 5 – 10 nm range. This interesting result is reasoned based on a balance between the relatively large valence band offset present at the STO/LFO interface coupled with the intrinsic semiconductive properties of LFO which produces an optimum range of 5-6 nm for OER catalysis.

## Experimental

*MBE Synthesis and Thin Film Characterization*

MBE synthesis was performed in an ultra-high vacuum system (Mantis Deposition, base pressure = $5x10^{-9}$ Torr) to achieve high purity in the resulting samples. *In situ* RHEED analysis was used to monitor the growth process. Commercially available (001)-oriented 0.7% Nb-doped $SrTiO_3$ (n-STO, MTI Crystal) 10 mm x 10 mm substrates were used. The small lattice mismatch (-0.6%) and conductive nature of n-STO makes it ideal as a conductive substrate suitable for catalysis measurements. Before loading into the growth chamber, n-STO substrates were sonicated for 10 minutes in acetone (ACS grade, VWR) and isopropanol (ACS grade, VWR) and dried with $N_2$ gas (Airgas, 99.999%) to produce a clean surface. La (ESPI, 99.9%) and Fe (Sigma Aldrich, 99.98%) flux were supplied through individual high temperature effusion cells. The cell temperatures for La and Fe were approximately 1550 °C and 1350 °C respectively. The oxygen flow was set to 0.2 sccm (Airgas, 99.999%) during the entire growth and an RF plasma source operated at 300 W was used to supply atomic oxygen. Pressure in the chamber during growth was $3x10^{-6}$ Torr. The *n*-STO substrate was heated to 700 °C in oxygen plasma before growth to remove hydrocarbons and recover a well-defined surface as observed by RHEED. A shuttered growth process was employed to synthesize the samples at 700 °C.[11,16] By programming individually controlled shutters, La and Fe fluxes were deposited alternately with a growth rate of ~80 seconds per unit cell. Samples were cooled in the presence of oxygen plasma until ~200 °C to promote full oxidation of the LFO film.

Epitaxially grown samples were then transferred to an appended X-ray photoelectron spectroscopy chamber (PHI 5400, refurbished by RBD Instruments). The XPS system is attached to the MBE system by a vacuum transfer line to preserve pristine surfaces for post-growth characterization.[17] A monochromatic Al K-alpha X-ray source was used for measurement. A low energy electron flood gun (neutralizer) was used for charge compensation. All the XPS peaks were adjusted by shifting the O 1*s* peak to 530 eV.[17] Following XPS, 10 mm x 10 mm n-STO/LFO samples were diced into (4) 5 mm x 5 mm samples for electrocatalysis and microscopy measurements. Post-growth high resolution X-ray diffraction (HRXRD) was performed using a Rigaku SmartLab system with a Ge(220)x2 incident beam monochromator and hybrid pixel area detector in 0D mode to obtain out-of-plane diffraction and X-ray reflectometry scans.

*OER Electrocatalysis*

*n*-STO/LFO (5 mm x 5 mm diced) samples were fabricated into electrodes using a rotating disk glassy carbon electrode (GC, Pine Instruments). Indium gallium eutectic (InGa, Ted Pella #495425) was used to form an electrical contact between the backside of the n-STO substrate and the GC electrode. A ring of silver paint (Sigma Aldrich) was placed around the InGa eutectic to serve as a binding agent between the n-STO substrate and the GC surface. Chemically inert epoxy (Loctite D609) was then used to cover the edges of the n-STO substrate and GC electrode, leaving only the LFO surface exposed to solution. Covering the entire edge of the substrate with epoxy was found to be critical to sealing any silver paint from exposure to solution. Failure to do this resulted in electrochemical behavior consistent with exposed silver, thus complicating electrocatalytic measurements. An image of the final fabricated electrode is shown in **Figure S1**.

Cyclic voltammetry (CV) experiments were performed with a Pine WaveDriver 20 bipotentiostat using a three-electrode setup. The working, reference, and counter electrodes for the electrochemical setup were n-STO/LFO, Hg/HgO (0.1 M NaOH, Pine Instruments), and platinum wire respectively. All measurements were performed under saturated $O_2$ conditions in water (18 MΩ, Millipore) with 0.1 M KOH electrolyte while rotating the working electrode at 2000 rpm to remove bubbles from the surface. All potentials were converted from Hg/HgO to RHE using the equation: $E_{RHE} = E_{app} + E_{Hg/HgO} + 0.059*pH$. $E_{Hg/HgO}$ is the potential of the reference electrode vs NHE measured by referencing to $[Fe(CN)_6]^{3-/4-}$ ($E^o = 0.36$ V vs NHE) before every experiment.[18] $E_{Hg/HgO}$ was routinely found to be 0.1 V vs NHE. The pH was measured before each experiment to be 12.5. CV experiments were performed by sweeping the potential at 20 mV $s^{-1}$ from 0.80 to 2.23 V vs RHE for 25 cycles to equilibrate the electrode surface. The anodic scan of the $25^{th}$ cycle was used for analysis of electrocatalytic performance. All potentials were iR compensated by measuring the solution resistance before each measurement.

Electrochemical impedance spectroscopy (EIS) was performed at designated applied potentials using a Gamry 1010E potentiostat with a 5 mV modulation voltage. The modulation frequency was scanned from 0.1 Hz to 100 kHz. EIS experiments were performed without rotation of the working electrode as this resulted in high noise in the low frequency regime. EIS data were fit with an equivalent circuit described in the text using AfterMath software (Pine Instruments).

*Scanning Transmission Electron Microscopy*

Cross-sectional scanning transmission electron microscopy (STEM) samples were prepared using a FEI Helios NanoLab DualBeam $Ga^+$ Focused Ion Beam (FIB) microscope with a standard lift out procedure. STEM high-angle annular dark field (STEM-HAADF) images were collected on a probe-corrected JEOL GrandARM-300F microscope operating at 300 kV, with a convergence semi-angle of 29.7 mrad and a collection angle range of 75–515 mrad. STEM electron energy loss spectroscopy (STEM-EELS) mapping was performed using a 0.25 eV $ch^{-1}$ dispersion for fine structure measurements, yielding an effective ~0.75 eV energy resolution, and using a 1 eV $ch^{-1}$ dispersion with a 4× energy binning in the dispersive direction for composition maps. Data was collected in the DualEELS mode to correct for energy drift and no denoising was applied.

## Results and Discussion

The RHEED pattern for the 6 nm LFO sample after cooling to ambient temperature is shown in **Figure 1a**. The sharp RHEED image taken along the [110] direction suggests the film is a single crystal with the perovskite structure. The absence of any modulation in the streaks in the RHEED image indicates a smooth surface with roughness comparable to the original substrate. Films show a c(2x1) surface reconstruction consistent with stoichiometric La-based perovskite films[19]. *In situ* XPS analysis was performed to determine the valence state of Fe and the film compositon. The Fe 2p XPS peak is shown in **Figure 1b**. The Fe 2p XPS region has two separate peaks corresponding to $2p_{1/2}$ and $2p_{3/2}$ due to spin multiplet splitting. An additional satellite peak close to the right shoulder of the $2p_{1/2}$ peak is an indication of the $Fe^{3+}$ valence state[20], while for $Fe^{2+}$ valence the satellite moves towards lower binding energies. Chemical composition analysis was done by comparing the area ratio under La 4d and Fe 2p peaks and were consistent with stoichiometric LFO. The LFO samples were also studied for film thickness and crystallinity using HRXRD. Out-of-plane HRXRD data is shown in **Figure 1c** and shows single phase films with peaks consistent with epitaxial films oriented along the [001] pseudocubic direction. All films exhibit a primary diffraction peak consistent with an out-of-plane lattice constant of ~3.93 Å, though precise determination is challenging for the thinnest samples. As the samples becomes thicker, finite thickness fringes are more distinct, indicating a smooth surface and interface with the *n*-STO substrate.

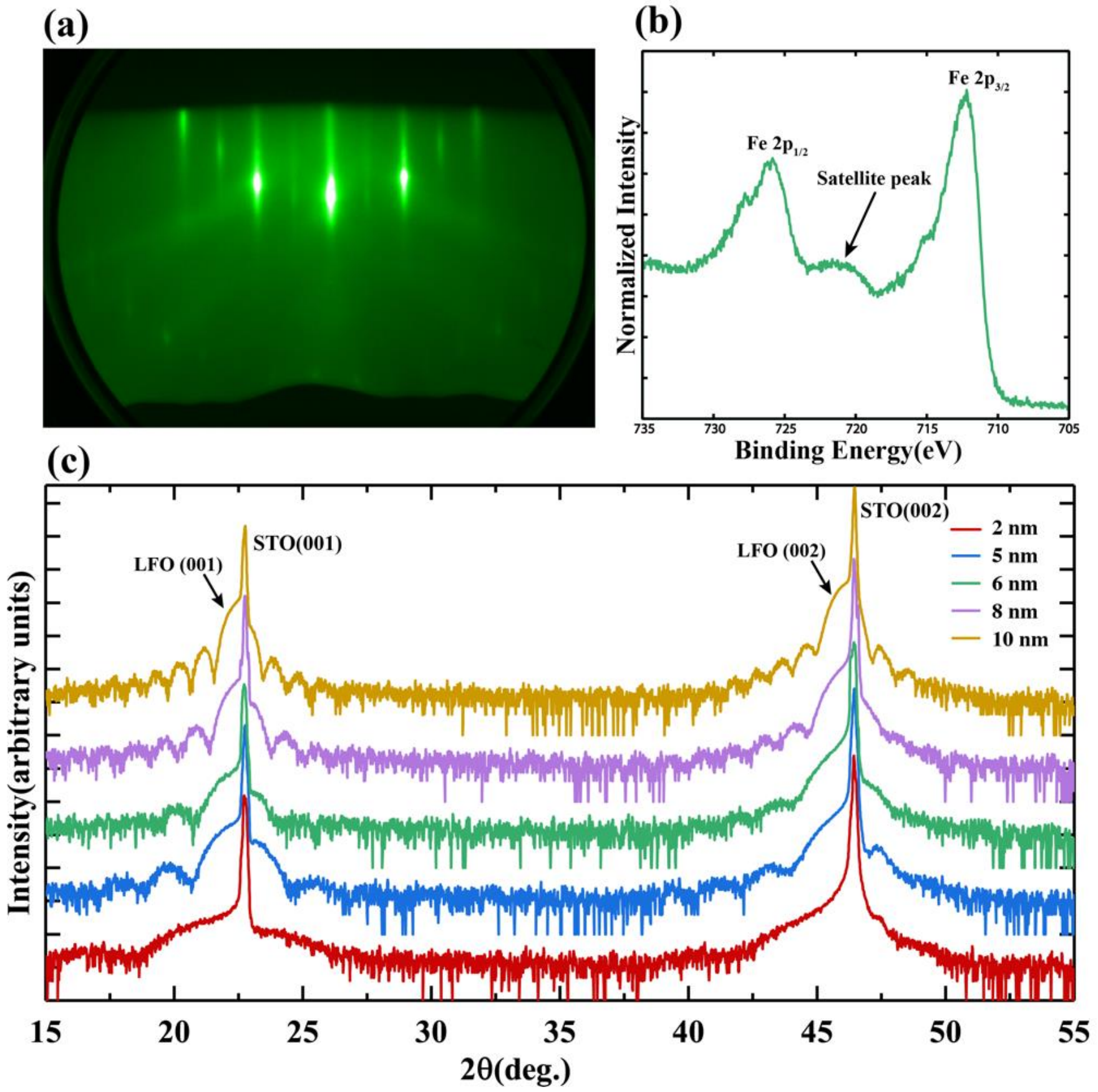

**Figure 1.** (a) RHEED pattern of a 6 nm LFO thin film sample along the [110] of the *n*-STO substrate. (b) Fe 2p XPS region for 6 nm LFO. The satellite on the right side of the of Fe $2p_{1/2}$ is a signature of the $Fe^{3+}$ valence state. (c) Out of plane HRXRD for n-STO/LFO samples of different thicknesses.

After electrode fabrication of the *n*-STO/LFO samples, cyclic voltammetry measurements were performed to examine the catalytic behavior of the LFO films for OER. **Figure 2a** shows CV data collected for LFO films of different thickness. These data were obtained following 25 cycles of the potential range (0.80 - 2.23 V vs RHE). The catalytic current density observed at 1.6 V vs RHE ($\eta_{OER}$ = 0.37 V) increased from 0.8 μA $cm^{-2}$ for 2 nm LFO to a maximum of 55.6 μA $cm^{-2}$ for 6 nm and then decreased to 2.7 μA $cm^{-2}$ for 10 nm (**Figure 2a inset**). Notably, the OER electrocatalysis observed for a bare *n*-STO substrate was significantly larger than the 2 nm LFO film, indicating that LFO was not etched from the surface and instead was passivating the catalysis of the n-STO surface.

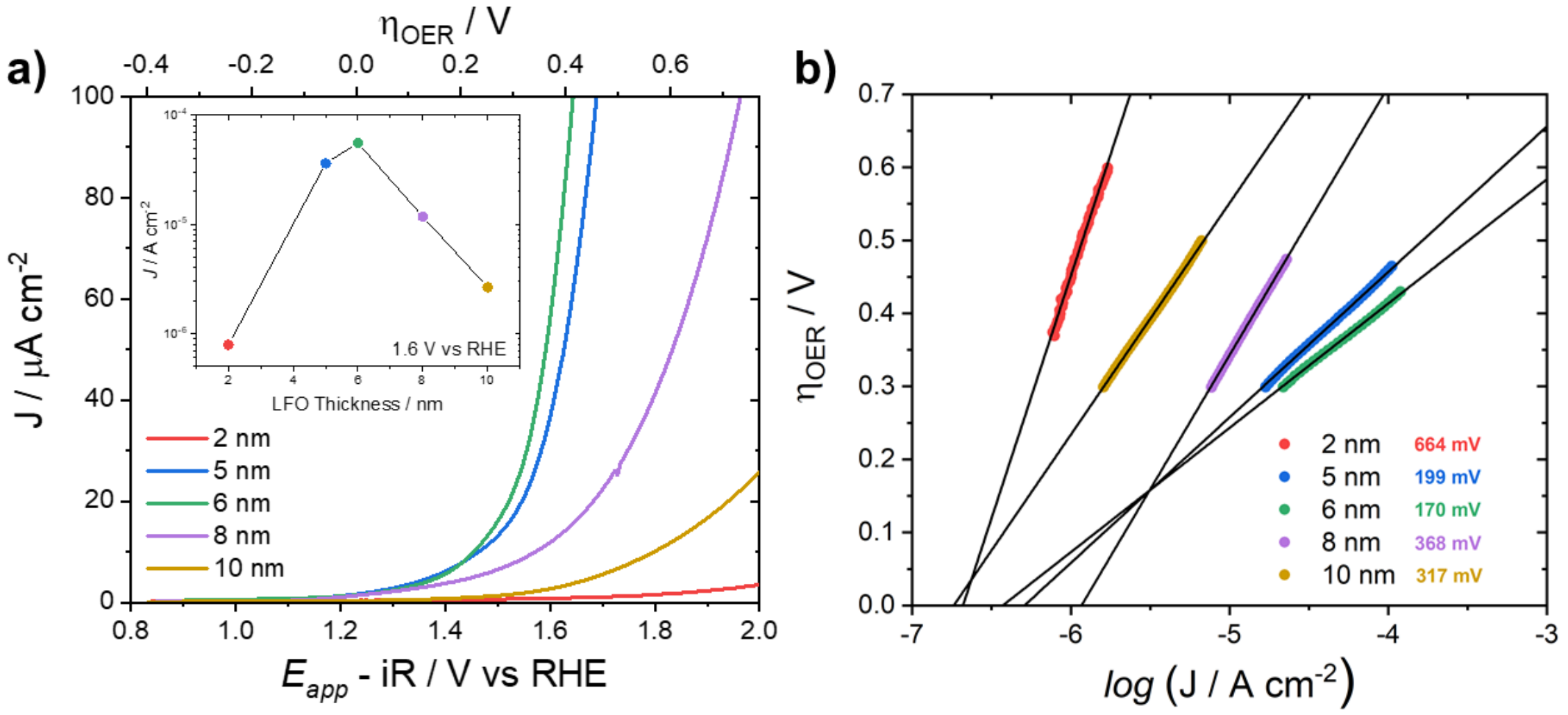

**Figure 2.** (a) Anodic scans obtained from CV for n-STO/LFO films of different thickness. Data collected in $O_2$ saturated 0.1 M KOH aqueous electrolyte at 20 mV $s^{-1}$ scan rate and 2000 rpm rotation. (Inset) Current density measured at 1.6 V vs RHE as a function of LFO thickness. (b) Tafel plot generated from CV measurements showing linear region between $\eta_{OER}$ = 0.3 – 0.6 V.

Tafel slopes were measured based on the linear dependences of log(J) vs $\eta_{OER}$ found in the region of 0.3-0.6 V (**Figure 2b**). These values were also found to change with LFO film thickness and yielded a minimum slope of 170 mV dec$^{-1}$ for the 6 nm LFO electrode. This value is large in comparison to Tafel slopes obtained for electrodes made from nanocrystalline LFO which have fallen in the range of 72 – 142 mV dec$^{-1}$.[21–24] The exchange current densities extrapolated to $\eta_{OER}$ = 0 V were found to be similar with each other and in the range of 0.2-1.3 μA cm$^{-2}$ (**Table 1**) The general trends observed for catalytic current and Tafel slopes with LFO film thickness point to a balance of competing factors for OER electrocatalysis with the best condition observed for a 6 nm film thickness.

**Table 1. Summary of OER Electrocatalytic Data**

| **LFO** **nm** | Tafel slope mV dec$^{-1}$ | $J_0$ μA cm$^{-2}$ | $J$ @ 1.6 V vs RHE μA cm$^{-2}$ |
|---|---|---|---|
| **2** | 664 | 0.2 | 0.8 |
| **5** | 199 | 0.5 | 36.3 |
| **6** | 170 | 0.4 | 55.6 |
| **8** | 368 | 1.3 | 11.8 |
| **10** | 317 | 0.2 | 2.7 |

In order to gain further understanding of the LFO thickness dependence, EIS was performed for each electrode for a selection of applied potentials. **Figure 3a-b** shows Nyquist and Bode-phase plots comparing 1.23, 1.63, and 2.23 V for 6 nm LFO where multiple features can be observed. As the applied potential was shifted positive from 1.23 V, the overall resistance through the LFO film decreased precipitously in the low frequency region while features in the mid-to-high frequencies were nearly constant. Accurate modeling of the data was obtained with the equivalent circuit model shown in **Figure 3e**. This model fundamentally represents three interfaces in series with Rs, a series resistor which accounts for resistance due to wires, clips, and electrolyte solution. Two of the interfaces were modeled as a parallel combination of resistors (R) and capacitors (i.e. constant phase elements (Q)) consistent with faradaic (i.e. charge-transfer across the interface) and non-faradaic (i.e. surface charging at the interface) current pathways through the circuit, respectively. These two interfaces represent features in the EIS data which occur at high and mid-frequencies and are assigned to the liquid/liquid interface of the reference electrode/electrolyte ($R_{ref}$, $Q_{ref}$) and the solid/solid interface of *n*-STO/LFO ($R_{STO}$, $Q_{STO}$), respectively. The third interface was modeled as a nested circuit, which described data in the low frequency region and has been used in the literature to explain the presence of surface electronic states.[25–27] This interface is assigned to the LFO/electrolyte interface where $Q_{LFO}$ represents the valence band capacitance of LFO, $Q_{SS}$ represents the capacitance of surface states, $R_{SS}$ represents charge-transfer resistance associated with catalysis driven by surface states, and $R_{LFO}$ represents charge-transfer associated with electrons moving between surface states and valence band states resulting in charge transport through the film. Nyquist and Bode-phase plots for all thickness and applied potentials are shown in **Figure S2** along with calculated resistance and capacitance values reported in **Table S1-S2**.

The necessity for the surface state model can be seen clearly in the Nyquist plot for 1.63 V (**Figure 3a**) where the confluence of two semicircle arcs is observed between $Z_{Re} = 10 – 100$ kΩ. This shows that a fourth interface is necessary to model the data. However, simulations with a simple linear series of four parallel RC circuits were unsuccessful. The nested model is much better at modeling data points between the two arcs in the Nyquist plot in the range of $Z_{Re} = 70$ kΩ. **Figure 3b** shows the corresponding Bode-phase plot where these features appear in the low frequency regime (< 100 Hz) while the mid frequency feature assigned to the *n*-STO/LFO interface is nearly constant around 1 kHz. A similar mid-frequency feature was observed by May et al. in a study of photoelectrocatalytic LFO films grown by pulsed layer deposition on STO substrates.[28] **Figure S3** shows individual simulations for each interface which highlight their contribution to the overall fit.

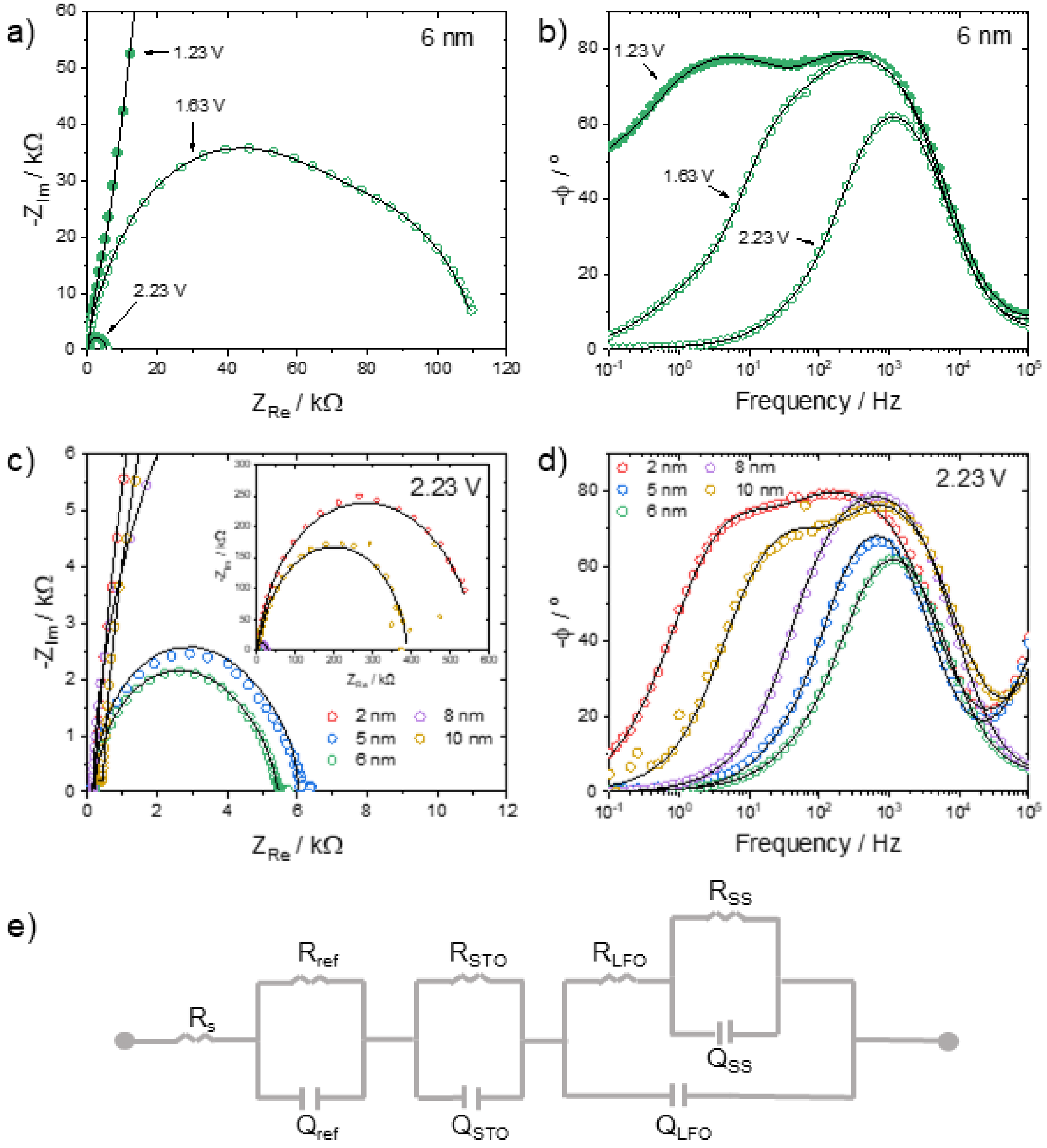

**Figure 3.** a) Nyquist and b) Bode-phase plots for 6 nm LFO measured at 1.23, 1.63, and 2.23 V vs RHE in 0.1 M KOH aqueous electrolyte (satd. $O_2$). c) Nyquist and d) Bode-phase plots for 2,

5, 6, 8, and 10 nm LFO films measured at 2.23 V vs RHE. e) Equivalent circuit used to fit experimental data and generate overlaid lines.

For all LFO thicknesses, the charge-transfer resistances in the low frequency region (i.e. $R_{SS}$ and $R_{LFO}$) decreased when the applied potential was shifted from 1.23 to 2.23 V vs RHE. This is consistent with an increase in OER catalysis through surface states and charge transport through the film to the n-STO substrate. **Figure 3c-d** shows Nyquist and Bode-phase plots for all thicknesses at 2.23 V where a similar thickness dependence observed from CV can be described for the resistance in the DC limit (i.e. $R_{DC} = Z_{Re}$ as $\omega \rightarrow 0$ Hz). Here, $R_{DC}$ was found to be minimized for the 5 nm (6.0 kΩ) and 6 nm (5.5 kΩ) conditions, those with the highest electrocatalytic current. $R_{DC}$ for 8 nm (28 kΩ) was the next smallest with 2 nm (560 kΩ) and 10 nm (390 kΩ) found to be much higher values. Comparing all resistances in the equivalent circuit, we can see that $R_{DC}$ is mostly comprised of $R_{STO}$ for the 5 and 6 nm samples when $R_{LFO} + R_{SS}$ is decreased to ~1 kΩ. In fact, $R_{STO}$ was nearly constant across all samples and applied potentials with an average of 5.4 ± 1.0 kΩ. This large resistance represents an inherent limitation of the electrode architecture due to the n-STO/LFO interface and is attributed to a significant valence band offset (VBO) due to band alignment between *n*-type Nb:STO and intrinsic/*p*-type LFO. This VBO has been estimated in the literature to be on the order of 2.0-2.2 eV and thus represents a large Schottky barrier for hole-transfer across the interface[11,14,28]. Importantly, the constant nature of $R_{STO}$ as a function of potential also indicates that the VBO is unchanged as a function of potential and that the applied potential is dropped primarily across the LFO/electrolyte interface.

EIS experiments were also used to produce Mott-Schottky plots for determination of the flat band potential ($E_{fb}$) at the LFO/electrolyte interface (**Figure S4**). For thicknesses of 5-10 nm, these plots revealed negative slopes indicative of *p*-type behavior with an average $E_{fb}$ = 1.38 ± 0.12 V vs RHE (**Table S3**). This value is similar to that reported by Wheeler and Choi for a nanocrystalline LFO film (1.45 V vs RHE)[29] but slightly larger than $E_{fb}$ reported by May et al. for PLD grown LFO films (1.0 V vs RHE)[28]. Interestingly, $E_{fb}$ was found to be -0.19 V vs RHE for 2 nm LFO and indicated n-type behavior based on the positive slope. Detailed XPS studies by Comes and Chambers have shown that band alignment at the *n*-STO/LFO interface can result in reduced Fe centers and thus *n*-type characteristics for such thin LFO layers (2 nm ~ 5 unit cells)[11]. May et al. also made a similar observation of *n*-type behavior for thin LFO deposited on STO based on photocatalytic OER studies.[28]

Given the large VBO at the *n*-STO/LFO interface and implication of surface state driven catalysis, questions arise regarding the structural and electronic variations throughout the depth of the LFO films. To examine these properties, STEM imaging and STEM-EELS measurements were performed on 6 nm samples with and without electrocatalysis experiments. **Figure 4** shows STEM high-angle annular dark-field (HAADF) imaging performed on these films for the pre-electrocatalytic and post-electrocatalytic conditions to investigate atomic changes at the surface induced by electrocatalysis. In this imaging mode, image intensity is proportional to atomic number ($Z^{\sim 1.7}$), so the heaviest La atom columns are brightest, and contrast can be directly interpreted to visualize film defects and potential mass loss as the sample surface. The as-grown film quality for the pre-electrocatalysis sample is excellent, with a smooth surface and no extended defects. Intermixing over several unit cells is present between the strontium and lanthanum atoms at the n-STO/LFO interface. Following electrocatalysis, we observed no substantial change in the n-STO/LFO interface and very little change at the LFO surface with only 1-2 u.c. steps present over large regions after treatment (**Figure S5**). Such small changes following exposure of the film surface to alkaline conditions and subjecting them to applied potentials reflects the stability of the MBE films during electrocatalysis. In addition, there is no evidence for a new phase in the treated sample, which would manifest in an interruption in the uniform crystalline structure if it formed during electrocatalysis.

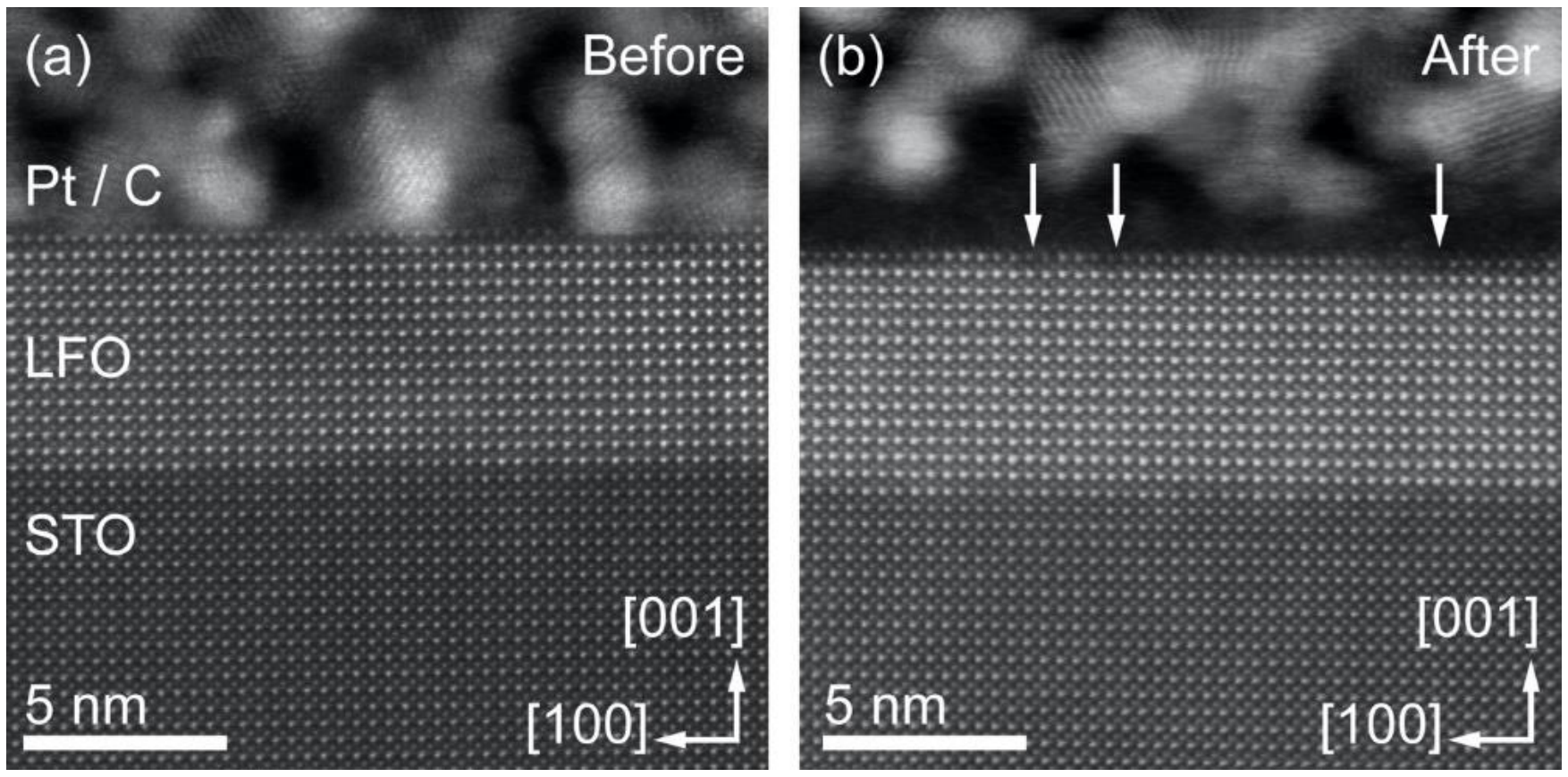

**Figure 4.** (a-b) Comparison of cross-sectional, high-magnification STEM-HAADF images of sample surface before and after treatment, respectively. Arrows indicate potential mass loss at the surface.

We also performed STEM-EELS measurements to explore chemical changes during cycling (**Figure 5**). A comparison of composition maps of the before and after samples is shown in **Figure 5a** and **Figure 5d**, respectively. Both maps are similar, with mixing on both the *A*- and *B*-site sublattices over ~1 u.c. at the interface. High-resolution O *K* and Fe $L_{2,3}$ edge spectra were collected from regions 1-3, corresponding to the interface, bulk, and surface of the samples, respectively. The O *K* edge shown in **Figure 5b** contains the expected three features: a pre-peak (labeled **a**), main peak (labeled **b**), and secondary peak (labeled **c**), which result from the hybridization of O 2*p* states with *B*-site 3*d*, La 5*d*, and *B*-site 4*sp* bands, respectively.[30] The overall line shape is comparable to prior work in the literature,[15] with a redistribution of the weight of peak **c** to slightly higher energy loss moving from the surface (region 3) to the interface (region 1). A similar line shape is observed in the sample after cycling, as shown in **Figure 5e**; however, in this case, there is a more pronounced pre-peak feature **a** and shift of main peak **b** to lower energy loss near the interface. Importantly, while the line shape varies only slightly throughout the untreated sample, it changes significantly near the surface of the treated sample, with a merging of pre-peak and main peak features. This finding points to possible electrocatalytically-induced changes in the oxygen environment and agrees with the microstructural changes observed in **Figure 4**. Furthermore, these changes may even be responsible for OER catalysis based on the surface state model discussed for EIS data above.

Inspection of the Fe $L_{2,3}$ edge (**Figure 5c** and **Figure 5f**) reveals the expected white-line doublet, whose edge onset is a known indicator of a change in valence state.[31] Moving from the surface to the interface, we observe a pronounced 1 eV shift of the $L_3$ edge to lower energy loss in both samples. This trend indicates clear reduction to a $Fe^{2+}$-like valence, in agreement the n-type Mott-Schottky behavior observed here for 2 nm LFO in addition to previous theoretical and experimental results that indicate the LFO conduction band is nearly degenerate with that of n-STO such that electrons will accumulate at the interface to reduce the Fe ions.[11,14,15]

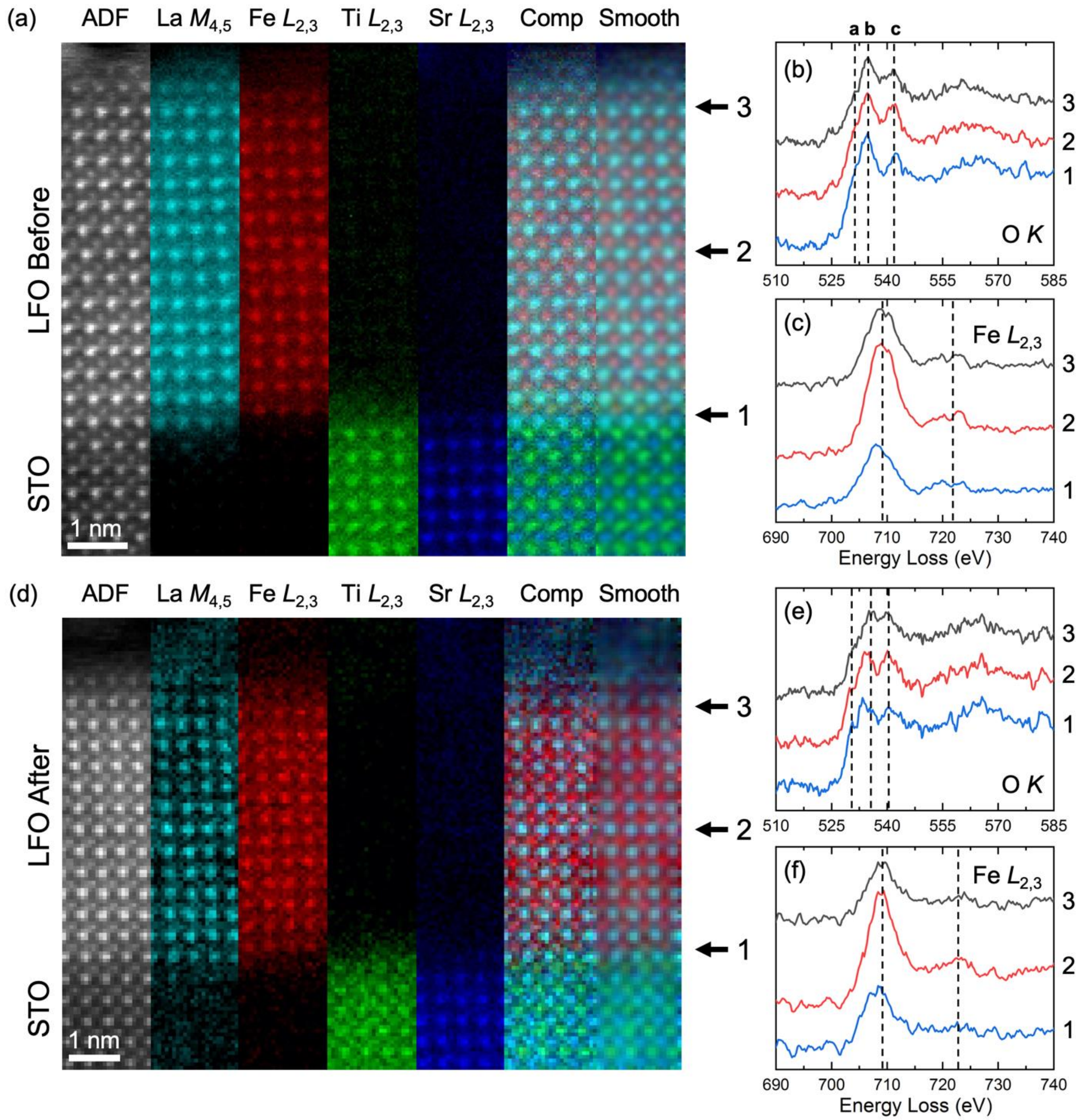

**Figure 5.** Cross-sectional STEM-ADF image, STEM-EELS composition maps, and extracted spectra for the O $K$ and Fe $L_{2,3}$ edges from the numbered regions for the before (a-c) and after (d-f) conditions. The extracted spectra were acquired from near the marked regions in a separate scan at higher dispersion. The dashed lines are added as a guide to the eye.

In order to interpret the spectral modifications observed for the O $K$ edge, a series of simulations computing the energy loss near edge spectra using the FDMNES code[32] has been carried out. The computational details of these simulations are provided in the supplementary information and a list of the structural deformations investigated is shown in **Table S4**. In addition to these simulations, the O *K* edge spectra of interstitial oxygen has been calculated as shown in **Figure S6f**. Interestingly, the spectral shape of the interstitial oxygen atom presents three features at energies very similar to the observed **a**, **b**, and **c** peaks. In particular, we note that the pre-peak **a** is more intense than the main peak **b**, itself being more intense than the secondary peak **c**. Based on a comparison between experimental spectra (**Figure S6e**), we note that the pre-edge feature in the O *K* edge is more distinct after electrocatalysis near the n-STO/LFO interface (region 1). Here, the O *K* edge spectra also presents a secondary peak **c** clearly less intense than the main peak **b**, which is not the case after electrocatalysis for spectra recorded from the middle and surface of the film. While this behavior could suggest the presence of interstitial oxygen atoms at the n-STO/LFO interface, caution should be taken as the spectra in region 1 after electrocatalysis also include a secondary peak **c**, which is more intense and shifted toward lower energies compared to the pre-electrocatalysis condition, as symbolized by the blue arrow in **Figure S6c**.

As shown in **Figure S6a-b**, similar spectral changes for secondary peak **c** are also observed in the middle and surface regions of the LFO film. As the changes observed in the secondary peak **c** are not reproduced by the calculated spectra for interstitial oxygen atoms alone, it suggests that structural deformations could also be involved. **Figure S7** summarizes the trends in calculated spectral changes (blue arrows) induced by structural deformations. While the filling between peak **b** and **c** after electrocatalysis (**Figure S6a**) is not indicative of the presence of interstitial oxygen, the trends in spectral changes (**Figure S7**) after electrocatalysis could involve several structural deformations such as a variation of the lattice angle $\gamma$, an expansion along the *b*-axis, and a reduction of the $FeO_6$ octahedral tilt, though no single model accounts for all of the variation in the spectra. Collectively these results confirm that the surface undergoes chemical and structural evolution during electrocatalysis but the exact chemical and structural transformation remains unclear.

Considering all electrochemical and structural characterization presented here, we propose an approximate band diagram shown in **Figure 6** to explain the thickness dependent OER electrocatalysis. The flat band potential, and thus the Fermi level ($E_F$), for the degenerately doped n-STO substrate (3.2 eV bandgap) was determined by Mott-Schottky measurements to be -0.59 V vs RHE (**Figure S7**) or 3.9 eV vs vacuum based on the conversion factor of 0 V vs RHE = 4.5 eV.

Fermi level equilibration between the *n*-STO substrate and LFO film ($E_{fb} = E_F = 1.38$ V vs RHE, 5.9 eV) results in a built-in potential (band bending) of ~1.8 eV at the interface. In order to reproduce a VBO ~2.2 eV reported in the literature for STO/LFO, the valence band edge for LFO was thus required to be 2.4 V vs RHE (6.9 eV). This estimate is more positive than previous estimates in the literature of 1.65 V vs RHE by Wheeler and Choi[29] and 1.10 V vs RHE by May et al.[28] In both of these reports, the valence band edge was assumed to be ~200 mV and ~100 mV below the measured flat band potential; however, no experimental evidence was given for these assumptions. These assumptions would indicate highly doped *p*-type LFO. In contrast, Comes and Chambers determined the difference to be much larger at ~1000 mV, indicating a more intrinsic material.[11] Here, we have used the same procedure for LFO growth as the Comes and Chambers report and thus use their band alignment determination as a basis for our model.

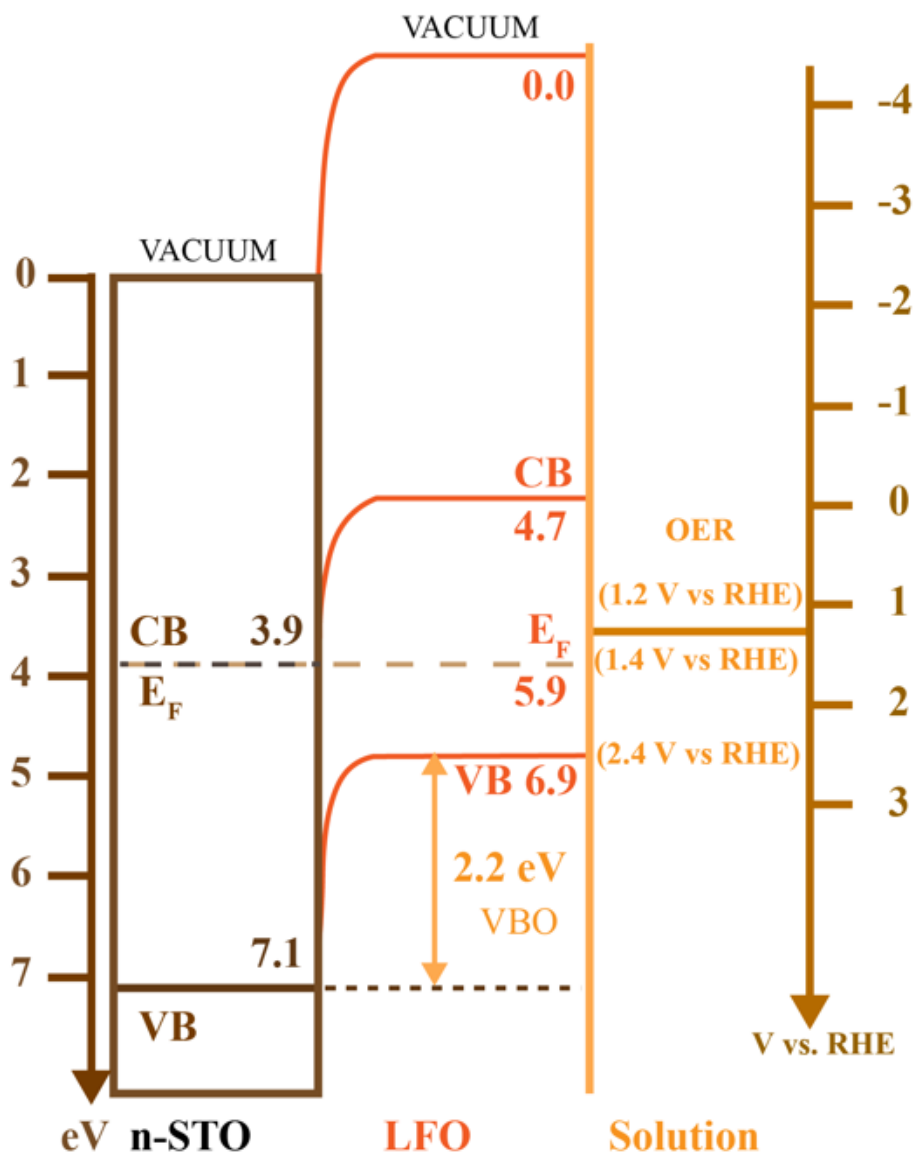

**Figure 6.** Band diagram for n-STO/LFO films immersed in aqueous electrolyte. Bandgaps of 3.2 and 2.2 eV were used for n-STO and LFO, respectively. Conversion from V vs RHE to eV was obtained using the conversion factor 0 V vs RHE = 4.5 eV.

Based on **Figure 6**, the thickness dependence for OER electrocatalysis with LFO can be explained based on the large VBO at the n-STO/LFO interface and the intrinsic nature of LFO. From previous literature, the VBO at the n-STO/LFO interface increases to a steady value of 2.0-2.2 eV within the first ~2.5 nm of deposited LFO. This means that the LFO/electrolyte interface for a 2 nm thick film displays more n-type character and results in poor catalysis. For LFO films >2.5 nm, the surface states which drive catalysis, most likely oxygen defects based on STEM-EELS results, are proximal to the n-STO/LFO interface and thus electrons have a higher probability of being extracted during OER electrocatalysis. Given the intrinsic nature of the LFO material and thus low concentration of free carriers in the valence band, this extraction likely relies on activation of electrons from the valence band to a low density of mid-gap electronic states, often called traps. This model represents a trap-state limited diffusion of charge through the LFO film and has been used to explain charge transport in nanocrystalline wide band gap oxides such as $TiO_2$[33–35]. As surface states are moved further from the *n*-STO/LFO interface, the probability of extraction decreases, resulting in lower observed current. Within our proposed circuit model, the $R_{LFO}$ term represents the activation of electrons from the valence band to mid-gap states. Notably, this term was found to be the limiting factor in determining the overall resistance at the LFO/electrolyte interface. This overall resistance can be taken as $R_{LFO} + R_{SS}$ where at the 2.23 V condition, $R_{LFO} >> R_{SS}$ for all LFO thicknesses. This indicates that overall current is more likely limited by activation and transport of carriers than by catalysis at the oxide surface. Consideration of bulk conductivity is thus a significant factor in the design of effective electrocatalysts, regardless of surface chemical reactivity. Furthermore, these results show that band engineering is critical for multilayer film catalysts. Future designs could be targeted to produce greater bulk conductivities with a surface layer that is band engineered for optimal catalytic performance. A recent work has begun to make inroads in this area, but further progress is needed.[36]

## Conclusions

In summary, we have examined the role that interfacial band alignment plays on OER electrocatalysis in $LaFeO_3$ using a series of thin films grown on *n*-$SrTiO_3$ by molecular beam epitaxy. We find that ultrathin films (~2 nm) are *n*-type due to the degeneracy of the STO and LFO conduction bands at the interface. As the thickness increases, band bending yields intrinsic or slightly *p*-type LFO with the Fermi level near mid-gap. For reasonably thin films this band alignment produces favorable electrocatalytic performance, but catalysis rapidly degrades with increasing thickness. Electrochemical impedance spectroscopy results indicate that surface reactivity is not the rate limiting factor in the catalytic process, but rather that catalysis is limited by a Schottky barrier at the n-STO/LFO interface coupled with slow electron transport in the bulk of the LFO film.

**Acknowledgments**

A.R.B, R.P., B.H.F., and R.B.C. acknowledge support from the National Science Foundation Division of Materials Research through grant NSF-DMR-1809847. HRXRD was performed with a Rigaku SmartLab instrument purchased with support from the National Science Foundation Major Research Instrumentation program through grant NSF-DMR-2018794. B.M., M.S., and S.R.S. acknowledge support from a Chemical Dynamics Initiative (CDi) Laboratory Directed Research and Development (LDRD) project at Pacific Northwest National Laboratory (PNNL). PNNL is a multiprogram national laboratory operated for the U.S. Department of Energy (DOE) by Battelle Memorial Institute under Contract No. DE-AC05-76RL0-1830. STEM sample preparation was performed in the Environmental Molecular Sciences Laboratory (EMSL), a national scientific user facility sponsored by the Department of Energy's Office of Biological and Environmental Research and located at PNNL. STEM imaging was performed in the Radiological Microscopy Suite (RMS), located in the Radiochemical Processing Laboratory (RPL) at PNNL.

## References

1 R. D. L. Smith, M. S. Prévot, R. D. Fagan, S. Trudel and C. P. Berlinguette, *J. Am. Chem. Soc.*, 2013, **135**, 11580–11586.
2 L. Trotochaud, J. K. Ranney, K. N. Williams and S. W. Boettcher, *J. Am. Chem. Soc.*, 2012, **134**, 17253–17261.
3 M. G. Walter, E. L. Warren, J. R. McKone, S. W. Boettcher, Q. Mi, E. A. Santori and N. S. Lewis, *Chem. Rev.*, 2010, **110**, 6446–6473.
4 B. M. Hunter, H. B. Gray and A. M. Müller, *Chem. Rev.*, 2016, **116**, 14120–14136.
5 J. Suntivich, K. J. May, H. A. Gasteiger, J. B. Goodenough and Y. Shao-Horn, *Science*, 2011, **334**, 1383–1385.
6 C. Baeumer, J. Li, Q. Lu, A. Y.-L. Liang, L. Jin, H. P. Martins, T. Duchoň, M. Glöß, S. M. Gericke, M. A. Wohlgemuth, M. Giesen, E. E. Penn, R. Dittmann, F. Gunkel, R. Waser, M. Bajdich, S. Nemšák, J. T. Mefford and W. C. Chueh, *Nat. Mater.*, 2021, **20**, 674–682.
7 K. A. Stoerzinger, R. Comes, S. R. Spurgeon, S. Thevuthasan, K. Ihm, E. J. Crumlin and S. A. Chambers, *J. Phys. Chem. Lett.*, 2017, **8**, 1038–1043.
8 J. R. Petrie, V. R. Cooper, J. W. Freeland, T. L. Meyer, Z. Zhang, D. A. Lutterman and H. N. Lee, *J. Am. Chem. Soc.*, 2016, **138**, 2488–2491.

9 L. Wang, K. A. Stoerzinger, L. Chang, X. Yin, Y. Li, C. S. Tang, E. Jia, M. E. Bowden, Z. Yang, A. Abdelsamie, L. You, R. Guo, J. Chen, A. Rusydi, J. Wang, S. A. Chambers and Y. Du, *ACS Appl. Mater. Interfaces*, 2019, **11**, 12941–12947.
10 L. Wang, K. A. Stoerzinger, L. Chang, J. Zhao, Y. Li, C. S. Tang, X. Yin, M. E. Bowden, Z. Yang, H. Guo, L. You, R. Guo, J. Wang, K. Ibrahim, J. Chen, A. Rusydi, J. Wang, S. A. Chambers and Y. Du, *Adv. Funct. Mater.*, 2018, **28**, 1803712.
11 R. Comes and S. Chambers, *Phys. Rev. Lett.*, 2016, **117**, 226802.
12 K. Nakamura, H. Mashiko, K. Yoshimatsu and A. Ohtomo, *Appl. Phys. Lett.*, 2016, **108**, 211605.
13 M. Nakamura, F. Kagawa, T. Tanigaki, H. S. Park, T. Matsuda, D. Shindo, Y. Tokura and M. Kawasaki, *Phys. Rev. Lett.*, 2016, **116**, 156801.
14 P. Xu, W. Han, P. M. Rice, J. Jeong, M. G. Samant, K. Mohseni, H. L. Meyerheim, S. Ostanin, I. V. Maznichenko, I. Mertig, E. K. U. Gross, A. Ernst and S. S. P. Parkin, *Adv. Mater.*, 2017, **29**, 1604447.
15 S. R. Spurgeon, P. V. Sushko, S. A. Chambers and R. B. Comes, *Phys. Rev. Mater.*, 2017, **1**, 063401.
16 J. H. Haeni, C. D. Theis, D. G. Schlom, W. Tian, X. Q. Pan, H. Chang, I. Takeuchi and X.-D. Xiang, *Appl. Phys. Lett.*, 2001, **78**, 3292–3294.
17 S. Thapa, R. Paudel, M. D. Blanchet, P. T. Gemperline and R. B. Comes, *J. Mater. Res.*, 2021, **36**, 26–51.
18 K. Gong, F. Xu, J. B. Grunewald, X. Ma, Y. Zhao, S. Gu and Y. Yan, *ACS Energy Lett.*, 2016, **1**, 89–93.
19 L. Qiao, K. H. L. Zhang, M. E. Bowden, T. Varga, V. Shutthanandan, R. Colby, Y. Du, B. Kabius, P. V. Sushko, M. D. Biegalski and S. A. Chambers, *Adv. Funct. Mater.*, 2013, **23**, 2953–2963.
20 S. A. Chambers, Y. J. Kim and Y. Gao, *Surf. Sci. Spectra*, 1998, **5**, 219–228.
21 S. She, J. Yu, W. Tang, Y. Zhu, Y. Chen, J. Sunarso, W. Zhou and Z. Shao, *ACS Appl. Mater. Interfaces*, 2018, **10**, 11715–11721.
22 X. Gao, Z. Sun, J. Ran, J. Li, J. Zhang and D. Gao, *Sci. Rep.*, 2020, **10**, 13395.
23 J. Dai, Y. Zhu, Y. Zhong, J. Miao, B. Lin, W. Zhou and Z. Shao, *Adv. Mater. Interfaces*, 2019, **6**, 1801317.
24 K. Lopez, G. Park, H.-J. Sun, J.-C. An, S. Eom and J. Shim, *J. Appl. Electrochem.*, 2015, **45**, 313–323.
25 A. R. C. Bredar, A. L. Chown, A. R. Burton and B. H. Farnum, *ACS Appl. Energy Mater.*, 2020, **3**, 66–98.
26 Q. Shi, S. Murcia-López, P. Tang, C. Flox, J. R. Morante, Z. Bian, H. Wang and T. Andreu, *ACS Catal.*, 2018, **8**, 3331–3342.
27 B. Klahr, S. Gimenez, F. Fabregat-Santiago, T. Hamann and J. Bisquert, *J. Am. Chem. Soc.*, 2012, **134**, 4294–4302.
28 K. J. May, D. P. Fenning, T. Ming, W. T. Hong, D. Lee, K. A. Stoerzinger, M. D. Biegalski, A. M. Kolpak and Y. Shao-Horn, *J. Phys. Chem. Lett.*, 2015, **6**, 977–985.
29 G. P. Wheeler and K.-S. Choi, *ACS Energy Lett.*, 2017, **2**, 2378–2382.
30 M. Varela, M. P. Oxley, W. Luo, J. Tao, M. Watanabe, A. R. Lupini, S. T. Pantelides and S. J. Pennycook, *Phys. Rev. B*, 2009, **79**, 085117.
31 H. Tan, J. Verbeeck, A. Abakumov and G. Van Tendeloo, *Ultramicroscopy*, 2012, **116**, 24–33.

32 O. Bunău and Y. Joly, *J. Phys. Condens. Matter*, 2009, **21**, 345501.
33 A. Hagfeldt and M. Graetzel, *Chem. Rev.*, 1995, **95**, 49–68.
34 J. Bisquert, *Phys. Rev. B*, 2008, **77**, 235203.
35 F. Fabregat-Santiago, I. Mora-Seró, G. Garcia-Belmonte and J. Bisquert, *J. Phys. Chem. B*, 2003, **107**, 758–768.
36 T. G. Yun, Y. Heo, H. Bin Bae and S.-Y. Chung, *Nat. Commun.*, 2021, **12**, 824.

**Thickness Dependent OER Electrocatalysis of Epitaxial $LaFeO_3$ Thin Films**

Andricus R. Burton[a†], Rajendra Paudel[b†], Bethany Matthews[c], Michel Sassi[d], Steven R. Spurgeon[c], Byron H. Farnum[a]*, Ryan B. Comes[b]*

[a]Department of Chemistry and Biochemistry, Auburn University, Auburn, AL 36849
[b]Department of Physics, Auburn University, Auburn, AL 36849
[c]Energy and Environment Directorate, Pacific Northwest National Laboratory, Richland, WA 99352
[d]Physical and Computational Sciences Directorate, Pacific Northwest National Laboratory, Richland, WA 99352

*corresponding authors: farnum@auburn.edu, ryan.comes@auburn.edu
†equal contribution

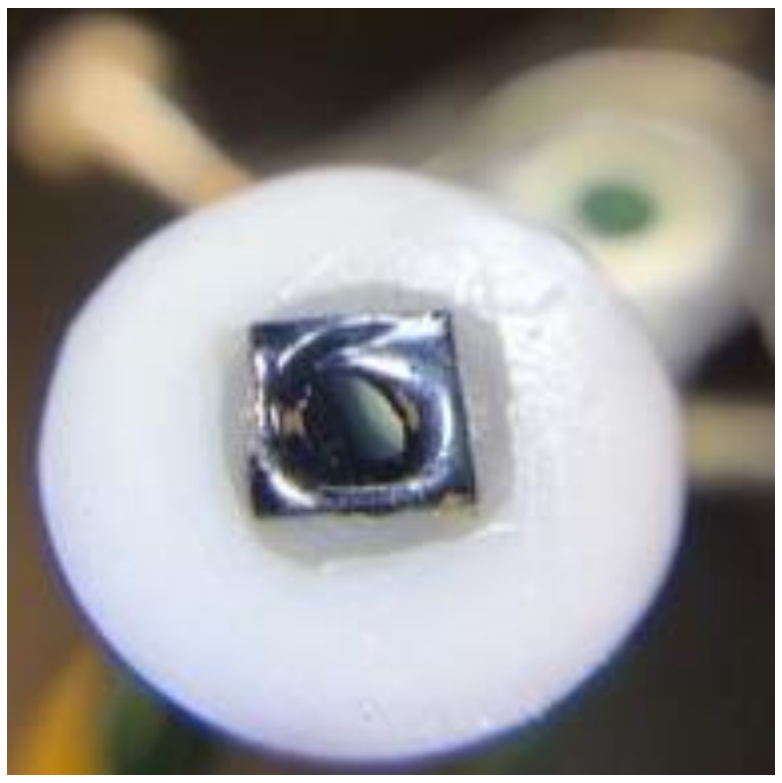

**Figure S1.** Fabricated n-STO/LFO on top of a GC rotating disk electrode. GC disk is completely covered by 5 mm x 5 mm n-STO substrate. Larger white area is a Teflon shaft.

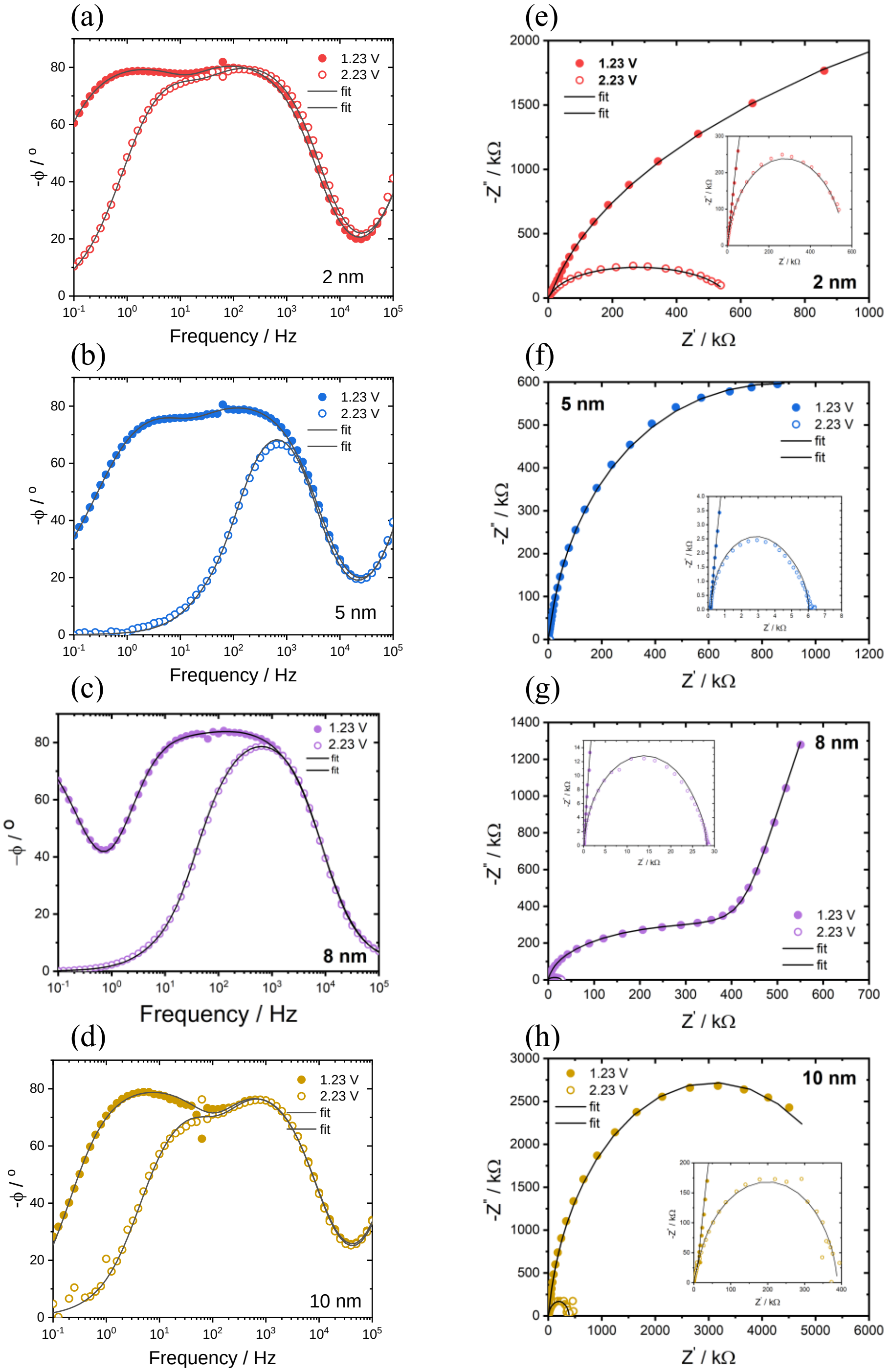

**Figure S2.** Bode-phase (a-d) and Nyquist (e-h) plots for 2, 5, 8, and 10 nm LFO films obtained from EIS experiments performed at 1.23 and 2.23 V. Overlaid fits were derived from the equivalent circuit model shown in **Figure 6e** of the main text.

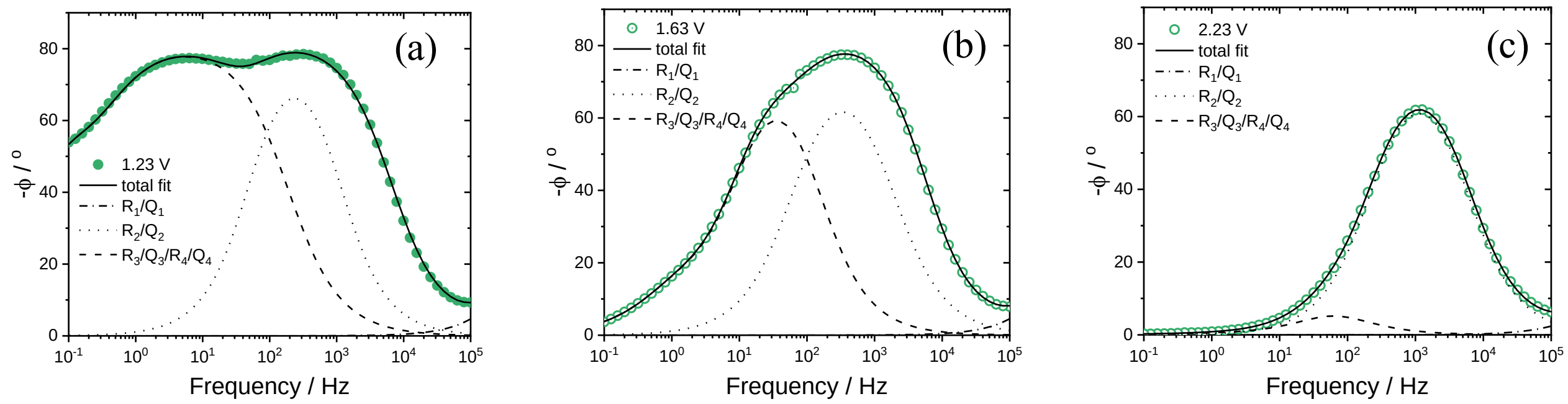

**Figure S3.** Bode-phase plots for 6 nm LFO at 1.23, 1.63, and 2.23 V. Solid line shows the full simulation according to the circuit model shown in **Figure 6e**. Dash-dotted, dotted, and dashed lines show individual simulations for different interfaces which contribute to the total fit. $R_1/Q_1$ = $R_{ref}/Q_{ref}$. $R_2/Q_2 = R_{STO}/Q_{STO}$. $R_3/Q_3/R_4/Q_4 = R_{LFO}/Q_{LFO}/R_{SS}/Q_{SS}$. As shown, $R_{STO}/Q_{STO}$ was mostly unchanged as a function of applied potential while $R_{LFO}/Q_{LFO}/R_{SS}/Q_{SS}$ decreased in magnitude as the potential was shifted positive.

**Table S1.** Summary of EIS Fitting Parameters for $E_{app}$ = 1.23 V vs RHE

| | **2 nm** | **5 nm** | **6 nm** | **8 nm** | **10 nm** |
|---|---|---|---|---|---|
| $Q_{ref}$ | 17 ± 82 nF $s^{\alpha-1}$ | 6.8 ± 10. nF $s^{\alpha-1}$ | 0.7 ± 3.2 nF $s^{\alpha-1}$ | 0.1 ± 6.6 nF $s^{\alpha-1}$ | 2.0 ± 5.4 nF $s^{\alpha-1}$ |
| $\alpha_{ref}$ | 0.92 ± 0.36 | 1.0 ± 0.1 | 1.0 ± 0.1 | 1.0 ± 1.8 | 1.0 ± 0.05 |
| $R_{ref}$ | 190 ± 23 Ω | 160 ± 21 Ω | 191 ± 26 Ω | 150 ± 24 Ω | 410 ± 61 |
| | | | | | |
| $Q_{STO}$ | 2.0 ± 1.9 μF $s^{\alpha-1}$ | 2.2 ± 1.8 μF $s^{\alpha-1}$ | 0.78 ± 0.73 μF $s^{\alpha-1}$ | 2.8 ± 3.6 μF $s^{\alpha-1}$ | 0.17 ± 0.15 μF $s^{\alpha-1}$ |
| $\alpha_{STO}$ | 0.99 ± 0.15 | 0.93 ± 0.11 | 1.0 ± 0.1 | 0.82 ± 0.10 | 0.98 ± 0.10 |
| $R_{STO}$ | 5.1 ± 0.0 kΩ | 5.4 ± 0.0 kΩ | 4.1 ± 0.0 kΩ | 5.0 ± 0.0 kΩ | 7.9 ± 0.0 kΩ |
| | | | | | |
| $Q_{LFO}$ | 0.56 ± 0.07 μF $s^{\alpha-1}$ | 0.62 ± 0.09 μF $s^{\alpha-1}$ | 0.36 ± 0.04 μF | 0.17 ± 0.03 μF $s^{\alpha-1}$ | 0.13 ± 0.02 μF $s^{\alpha-1}$ |
| $\alpha_{LFO}$ | 0.92 ± 0.02 | 0.93 ± 0.03 | 0.91 ± 0.02 | 0.99 ± 0.03 | 0.92 ± 0.04 |
| $R_{LFO}$ | 5.8 ± 0.0 MΩ | 1.0 ± 0.0 MΩ | 2.7 ± 0.0 MΩ | 0.55 ± 0.0 MΩ | 5.5 ± 0.0 MΩ |
| | | | | | |
| $Q_{SS}$ | 2.1 ± 9.7 μF $s^{\alpha-1}$ | 2.4 ± 2.3 μF $s^{\alpha-1}$ | 0.58 ± 0.42 μF | 1.1 ± 0.2 μF $s^{\alpha-1}$ | 0.032 ± 1.4 μF $s^{\alpha-1}$ |
| $\alpha_{SS}$ | 0.5 ± 4.0 | 0.64 ± 0.71 | 0.91 ± 0.44 | 0.94 ± 0.09 | 1.0 ± 8.6 |
| $R_{SS}$ | 21 ± 0.0 MΩ | 1.0 ± 0.0 MΩ | 4.6 ± 0.0 MΩ | 71 ± 0.0 MΩ | 0.69 ± 0.0 MΩ |

$R_s$ was set to 1 Ω for all fits; α is unitless; error expressed as standard error from the fitting analysis; an error of 0.0 indicates that the error was smaller than the significant figures of the mean value

**Table S2.** Summary of EIS Fitting Parameters for $E_{app}$ = 2.23 V vs RHE

| | **2 nm** | **5 nm** | **6 nm** | **8 nm** | **10 nm** |
|---|---|---|---|---|---|
| $Q_{ref}$ | 19 ± 114 nF $s^{\alpha-1}$ | 13 ± 31 nF $s^{\alpha-1}$ | 1.0 ± 14.1 nF $s^{\alpha-1}$ | 4.8 ± 367.2 nF $s^{\alpha-1}$ | 1.9 ± 5.3 nF $s^{\alpha-1}$ |
| $\alpha_{ref}$ | 0.91 ± 0.45 | 0.95 ± 0.18 | 0.91 ± 0.80 | 0.80 ± 5.47 | 1.0 ± 0.0 |
| $R_{ref}$ | 190 ± 29 Ω | 170 ± 23 Ω | 210 ± 33 Ω | 150 ± 35 Ω | 420 ± 63 Ω |
| | | | | | |
| $Q_{STO}$ | 1.6 ± 1.4 μF $s^{\alpha-1}$ | 0.31 ± 0.18 μF $s^{\alpha-1}$ | 0.25 ± 0.15 μF $s^{\alpha-1}$ | 0.78 ± 1.5 μF $s^{\alpha-1}$ | 0.16 ± 0.17 μF $s^{\alpha-1}$ |
| $\alpha_{STO}$ | 0.92 ± 0.10 | 0.98 ± 0.06 | 0.94 ± 0.06 | 0.92 ± 0.18 | 1.0 ± 0.1 |
| $R_{STO}$ | 5.5 ± 0.0 kΩ | 4.8 ± 0.0 kΩ | 4.1 ± 0.2 kΩ | 4.2 ± 0.6 kΩ | 5.9 ± 0.0 kΩ |
| | | | | | |
| $Q_{LFO}$ | 0.41 ± 0.08 μF $s^{\alpha-1}$ | 5.5 ± 37.4 μF $s^{\alpha-1}$ | 5.6 ± 17.7 μF | 0.19 ± 0.13 μF $s^{\alpha-1}$ | 0.12 ± 0.34 μF |
| $\alpha_{LFO}$ | 0.95 ± 0.04 | 1.0 ± 1.3 | 0.87 ± 0.56 | 1.0 ± 0.1 | 0.93 ± 0.05 |
| $R_{LFO}$ | 4.9 ± 0.0 kΩ | 0.96 ± 0.13 kΩ | 1.0 ± 0.4 kΩ | 23 ± 0.6 kΩ | 360 ± 0.0 kΩ |
| | | | | | |
| $Q_{SS}$ | 5.8 ± 16.2 μF $s^{\alpha-1}$ | 49 ± 7,922 μF $s^{\alpha-1}$ | 4,600 ± 29,400 μF | 18 ± 44 μF $s^{\alpha-1}$ | 1.1 ± 14.3 μF |
| $\alpha_{SS}$ | 0.87 ± 1.27 | 1.0 ± 25.4 | 0.64 ± 1.3 | 1.0 ± 0.5 | 1.0 ± 3.8 |
| $R_{SS}$ | 66 ± 0.0 kΩ | 0.07 ± 0.13 kΩ | 0.09 ± 0.37 kΩ | 1.3 ± 0.6 kΩ | 20. ± 1.8 kΩ |

$R_s$ was set to 1 Ω for all fits; α is unitless; error expressed as standard error from the fitting analysis; an error of 0.0 indicates that the error was smaller than the significant figures of the mean value

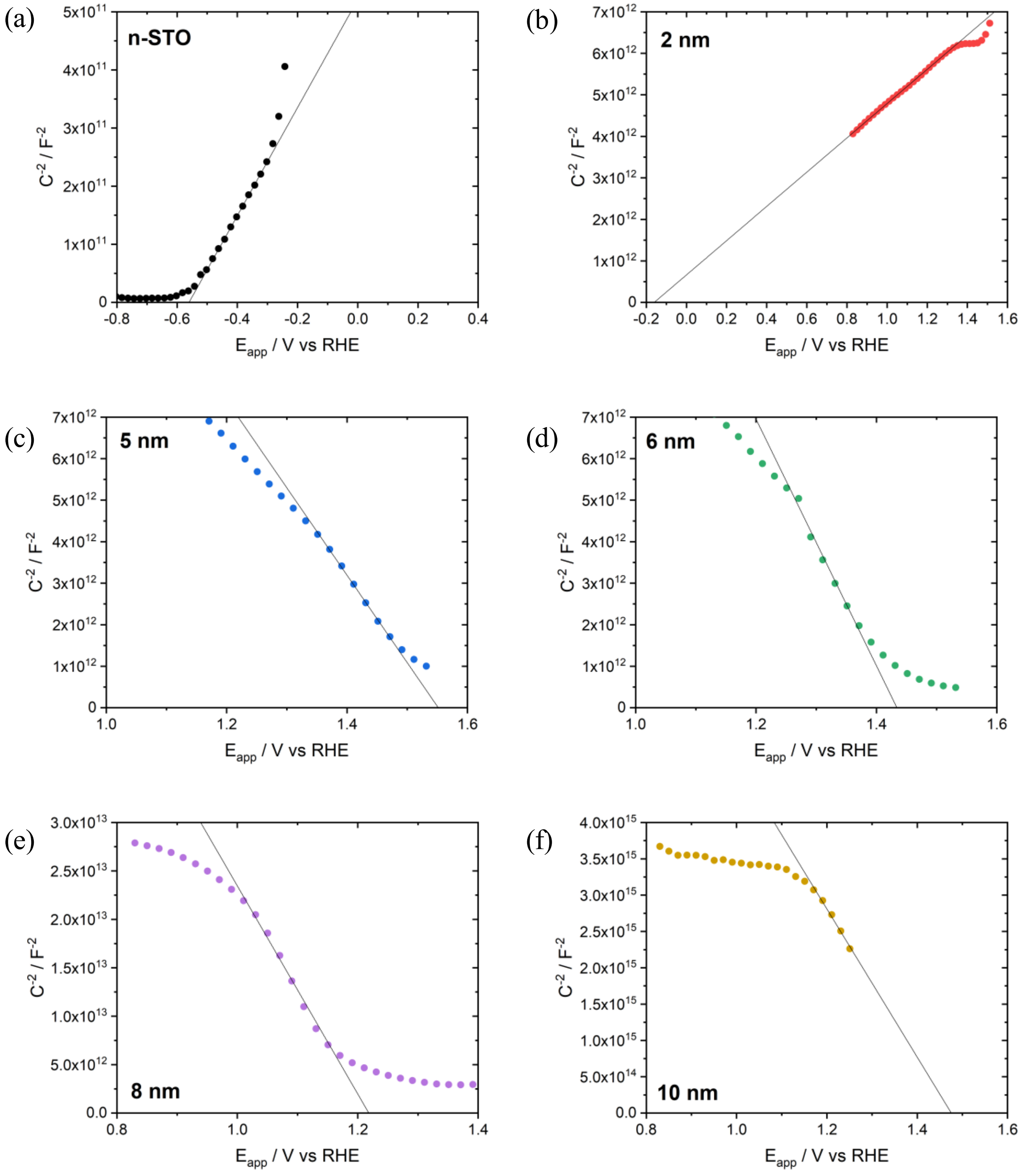

**Figure S4.** Mott-Schottky plots obtained at 0.5 Hz for (a-f) 0 (n-STO), 2, 5, 6, 8, and 10 nm LFO films.

**Table S3. Summary of Mott-Schottky data**

| **LFO / nm** | $E_{fb}$ / V vs RHE[a] | VBO / eV[b] |
|---|---|---|
| **2** | -0.19 | 0.61 |
| **5** | 1.52 | 2.32 |
| **6** | 1.40 | 2.20 |
| **8** | 1.19 | 1.99 |
| **10** | 1.43 | 2.23 |

[a]$E_{fb}$ calculated from the Mott-Schottky equation; [b]VBO = $E_{fb}$ – $E_{vb}$(LFO) + $E_g$(STO); $E_{vb}$ (LFO) = 2.4 V vs RHE; $E_g$(STO) = 3.2 eV

**Mott-Schottky Equation**

$$\frac{1}{C^2} = \frac{2}{\varepsilon\varepsilon_o A^2 q N_A}\left(E_{app} - E_{fb} - k_B T/q\right)$$

C = interfacial capacitance (F)
ε = relative permittivity
$\varepsilon_o$ = vacuum permittivity (8.85 x $10^{-12}$ F $m^{-1}$)
A = electrode surface area ($cm^2$)
Q = fundamental charge (1.602 x $10^{-19}$ C)
$N_A$ = density of acceptors ($cm^{-3}$)
$E_{app}$ = applied potential (V vs RHE)
$E_{fb}$ = flat band potential (V vs RHE)
$k_B$ = Boltzmann constant (1.38 x $10^{-23}$ J $K^{-1}$)
T = temperature (298 K)

Based on the Mott-Schottky equation, $E_{fb}$ was estimated by subtracting $k_B$T/q (= 0.0256 V) from the x-intercept of the Mott-Schottky plots.

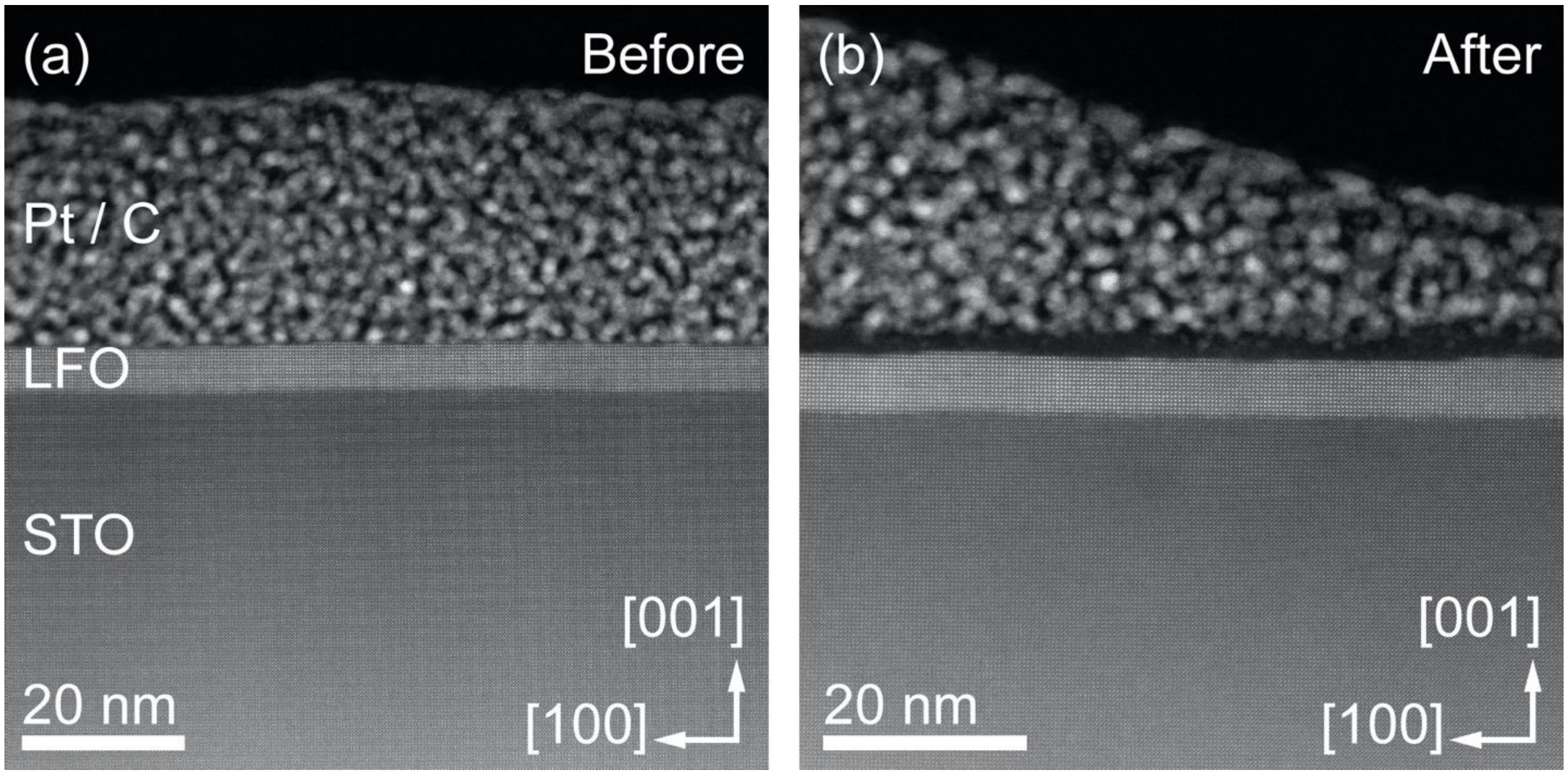

**Figure S5.** (a-b) Comparison of cross-sectional, low-resolution STEM-HAADF images of sample surface before and after treatment, respectively.

**Computational details for STEM-EELS data**

The simulations of O *K* edge XANES spectra were performed with the FDMNES code[1] and used the experimental orthorhombic $LaFeO_3$ structure of Selbach *et al.*[2] Although EELS and XANES are not strictly equivalent techniques, they probe the same electronic states. Therefore, a comparison between experimental and theoretical spectra across the two techniques provide invaluable insight and are commonly used to rationalize observed trends and fine structure features in oxides.[3] In FDMNES, the final excited state is obtained by solving a Schrödinger-like equation through the Greens formalism, within the limit of the muffin-tin approximation. The potentials and Fermi energy were determined self-consistently using a radii of 7 Å. Similar radii were used for the calculations of the spectra. Real Hedin-Lundquist potentials[4] were used to model the exchange-correlation. Dipoles, quadrupoles, core-hole and spin-orbit contributions were taken into account. A Hubbard correction of 4 eV has been applied to the localized valence orbitals of the Fe species.

In the case of oxygen interstitial, the atomic coordinates have been relaxed by density functional theory simulations (DFT) prior to calculating the XANES with FMNDES, while fixing the lattice parameters to their experimental values. The DFT calculations were performed with the VASP package[5] and used the PBEsol exchange-correlation functional.[6] The calculations of interstitial oxygen in orthorhombic $LaFeO_3$ used a 2×2×2 supercell of the experimental structure from Selbach *et al.* The cutoff energy for the plane wave basis set was fixed to 550 eV and a Monkhorst-Pack[7] *k*-points mesh of 2×2×2 for the sampling of the Brillouin zone was used. The total energy was converged to $10^{-5}$ eV/cell and the force components on the atoms were relaxed to below $10^{-4}$ eV/Å. Spin-polarization were used and the GGA+U method, as described by Dudarev,[8] was applied for the Fe atoms to correct the description of the Coulomb repulsion of the 3d electrons in standard GGA. The Hubbard parameter, *U*, describing the Coulomb interaction, was fixed to 5 eV, while the screened exchange energy, *J*, was fixed to 1 eV.

| | Deformation | a (Å) | b (Å) | c (Å) | Fe—O—Fe (°) (in-plane) | Fe—O—Fe (°) (out-of-plane) | dFe—O (Å) (in-plane) | dFe—O (Å) (out-of-plane) |
|---|---|---|---|---|---|---|---|---|
| | **Reference (Expt.)** | 5.554 | 5.566 | 7.853 | 154.81 | 157.37 | 1.975/2.053 | 2.002 |
| Independent variation of lattice parameters | **a expansion (5%)** | 5.832 | 5.566 | 7.853 | 155.30 | 156.30 | 2.033/2.093 | 2.006 |
| | **a contraction (5%)** | 5.276 | 5.566 | 7.853 | 154.31 | 158.46 | 1.919/2.014 | 1.999 |
| | **b expansion (5%)** | 5.544 | 5.844 | 7.853 | 155.27 | 157.39 | 2.012/2.111 | 2.002 |
| | **b contraction (5%)** | 5.544 | 5.288 | 7.853 | 154.32 | 157.40 | 1.994/1.938 | 2.002 |
| | **c expansion (5%)** | 5.544 | 5.566 | 8.346 | 153.56 | 158.72 | 1.978/2.056 | 2.123 |
| | **c contraction (5%)** | 5.554 | 5.566 | 7.460 | 155.78 | 156.21 | 2.049/1.972 | 1.906 |
| Deformations based on LFO/SFO lattice mismatch | **LFO (2a,2b)/STO (3a,3a)** | 5.858 | 5.858 | 7.853 | 155.79 | 156.17 | 2.078/2.159 | 2.006 |
| | **LFO (-2a,-2b)/STO (3a,3a)** | 5.263 | 5.263 | 7.853 | 153.70 | 158.53 | 1.947/1.875 | 1.998 |
| | **LFO (2b,1c)/STO (3a,2a)** | 5.554 | 5.858 | 7.810 | 155.42 | 157.23 | 2.015/2.115 | 1.992 |
| | **LFO (-2b,-1c)/STO (3a,2a)** | 5.554 | 5.263 | 7.896 | 154.17 | 157.52 | 1.990/1.935 | 2.013 |
| Variation of octahedral tilt | **No octahedral tilt** | 5.554 | 5.566 | 7.853 | 180.00 | 180.00 | 1.966/1.966 | 1.963 |
| | **Increase In-plane Oct. tilt** | 5.554 | 5.566 | 7.853 | 156.36 | 157.37 | 1.893/2.124 | 2.002 |
| | **Reduction In-plane Oct. tilt** | 5.554 | 5.566 | 7.853 | 151.25 | 157.37 | 1.954/2.105 | 2.002 |
| Variation of lattice angle | **Latt. Angle α (95°)** | 5.554 | 5.566 | 7.853 | 154.23/154.23 | 157.37 | 2.079/1.954 | 1.997 |
| | **Latt. Angle α (85°)** | 5.554 | 5.566 | 7.853 | 155.46/155.46 | 157.37 | 2.027/1.997 | 2.007 |
| | **Latt. Angle β (95°)** | 5.554 | 5.566 | 7.853 | 154.25/155.44 | 157.45 | 2.031/1.949 | 2.035 |
| | **Latt. Angle β (85°)** | 5.554 | 5.566 | 7.853 | 155.44/154.25 | 157.45 | 2.074/2.001 | 1.969 |
| | **Latt. Angle γ (95°)** | 5.554 | 5.566 | 7.853 | 156.01/153.45 | 157.11 | 2.136/1.893 | 2.003 |
| | **Latt. Angle γ (85°)** | 5.554 | 5.566 | 7.853 | 153.45/156.01 | 157.65 | 1.967/2.055 | 2.001 |

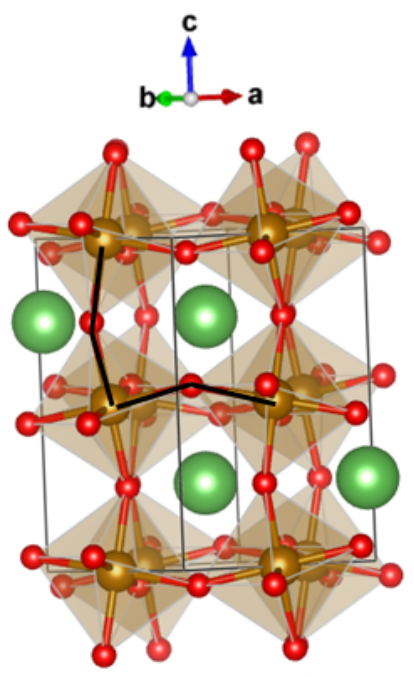

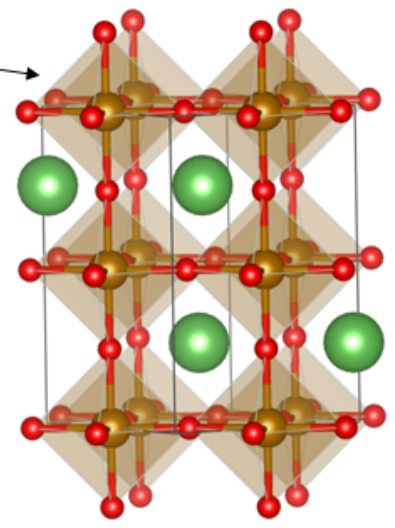

**Table S4**. List of the structural deformations investigated. Atoms represented by green, orange, and red balls are La, Fe, and O species respectively.

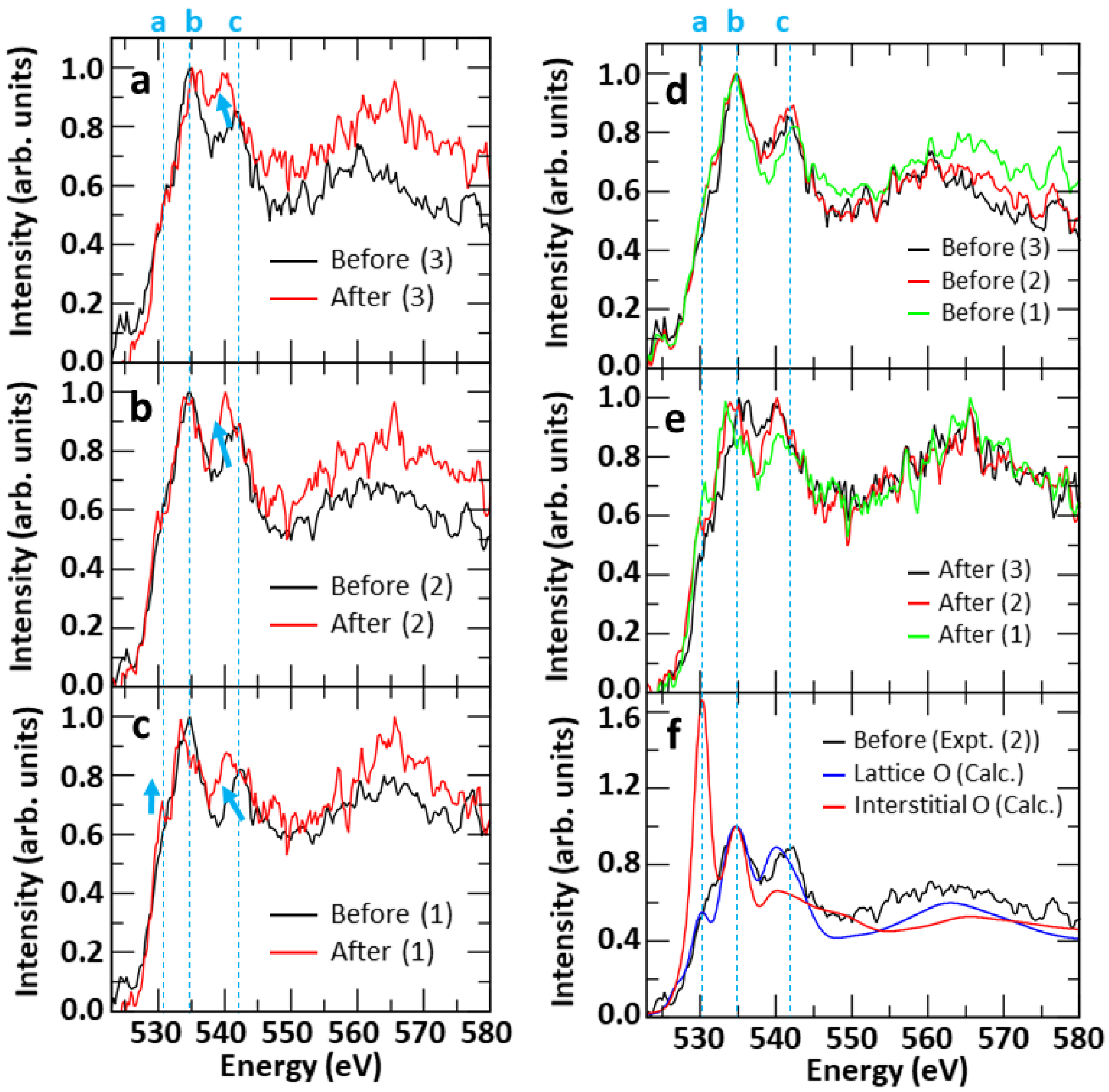

**Figure S6**. Comparison of experimental O *K* edge spectra before and after electrocatalysis for the different regions (1), (2), and (3) corresponding the film-substrate interface, middle, and film's surface respectively (a-e). (f) Comparison between experimental (2) and calculated spectra for lattice oxygen and interstitial oxygen.

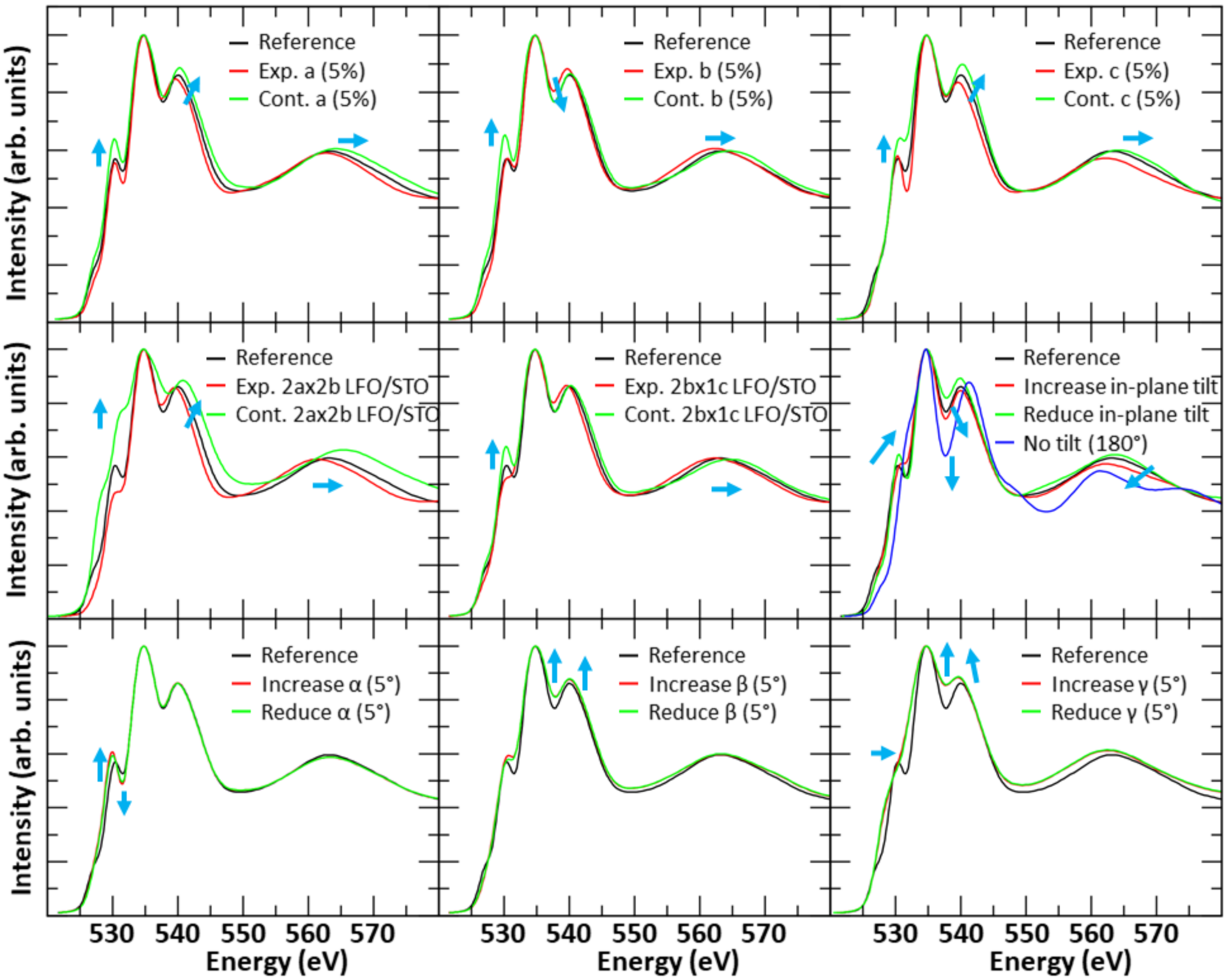

**Figure S7**. Comparison of the spectral changes as induced by various structural deformations listed in **Table S3**.

**References**

1. Bunău, O.; Joly, Y. Self-consistent aspects of x-ray absorption calculations. *Journal of Physics: Condensed Matter* **2009**, *21*, 345501.
2. Selbach, S.M.; Tolchard, J.R.; Fossdal, A.; Grande, T. Non-linear thermal evolution of the crystal structure and phase transitions of $LaFeO_3$ investigated by high temperature X-ray diffraction. *Journal of Solid State Chemistry* **2012**, *196*, 249-254.
3. Steven R. Spurgeon, Prasanna V. Balachandran, Despoina M. Kepaptsoglou, Anoop R. Damodaran, J. Karthik, Siamak Nejati, Lewys Jones, Haile Ambaye, Valeria Lauter, Quentin M. Ramasse, Kenneth K.S. Lau, Lane W. Martin, James M. Rondinelli, and Mitra L. Taheri. Polarization screening-induced magnetic phase gradients at complex oxide interfaces. Nature Communications 2015, 6, 6735.
4. Hedin, L.; Lundqvist, B.I. Explicit local exchange-correlation potentials *J. Phys. C: Solid State Phys.* **1971**, *4*, 2064.
5. Kresse, G.; Furthmuller, J. Efficient iterative schemes for ab initio total-energy calculations using a plane-wave basis set. *Phys. Rev. B* **1996**, *54*, 11169-11186.
6. Perdew, J.P.; Ruzsinszky, A.; Csonka, G.I.; Vydrov, O.A.; Scuseria, G.E.; Constantin, L.A.; Zhou, X.; Burke, K. Restoring the Density-Gradient Expansion for Exchange in Solids and Surfaces. *Physical Review Letters* **2008**, *100*, 136406.
7. Monkhorst, H.J.; Pack, J.D. Special points for Brillouin-zone integrations. *Physical Review B* **1976**, *13*, 5188-5192.
8. Dudarev, S.L.; Botton, G.A.; Savrasov, S.Y.; Szotek, Z.; Temmerman, W.M.; Sutton, A.P. Electronic structure and elastic properties of strongly correlated metal oxides from first principles: LSDA+U, SIC-LSDA and EELS study of $UO_2$ and NiO. *Phys. Status Solidi.* **1998**, *166*, 429-443.